\documentclass{IEEEtran}
\usepackage{cite}
\usepackage{caption,physics}
\usepackage{subcaption}
\usepackage{amsmath,amssymb,amsfonts,amsthm}
\usepackage{algorithm}
\usepackage{algorithmicx}
\usepackage{algcompatible}
\usepackage{algpseudocode}
\usepackage{dsfont}
\usepackage{booktabs}
\usepackage{enumitem}

\usepackage{mathabx}
\usepackage{graphicx,xcolor}
\usepackage{subcaption}
\usepackage{multirow}
\usepackage{textcomp}
\usepackage{bbm}
\definecolor{darkgreen}{rgb}{0,0.5,0}

\usepackage{subfig}
\usepackage{physics}

\newlist{myitem}{itemize}{1}
\setlist[myitem]{           
    label=\textbullet,       
    leftmargin=1.5em,        
    labelsep=1em,          
    itemsep=0pt,             
    topsep=0pt,
    align=parleft           
}

\def\BibTeX{{\rm B\kern-.05em{\sc i\kern-.025em b}\kern-.08em
    T\kern-.1667em\lower.7ex\hbox{E}\kern-.125emX}}

\newcommand{\im}{\mathbbm{i}}

\begin{document}
\title{Simplest Nontrivial Maxwellian Random Field Models for Stochastic LoS MIMO Using the Dyadic Green's Function}

\author{Lumeng Xu and Said Mikki
\thanks{Lumeng Xu and Said Mikki are with Zhejiang University/University of Illinois at Urbana-Champaign (ZJU-UIUC) Institute, Haining, Zhejiang University, Zhejiang, China. The corresponding author is Said Mikki (smikki@illinois.edu). }

       }

\maketitle

\begin{abstract}
This letter introduces a novel, full-wave, physics-
compliant stochastic dyadic Green's function (SDGF) framework for modeling electromagnetic (EM) multiple-input-multiple-output (MIMO) channels under wavenumber uncertainty.
Unlike conventional phenomenological fading models, the proposed
approach provides what appear to be the simplest exact random field models of electromagnetic line-of-sight (LoS) propagation
that are also exact solutions of Maxwell's equations. Hence,
we dub them Maxwellian random field theoretic models. These
physically consistent stochastic models—an analytically tractable
wavenumber Gaussian model and a more general stochastic plane wave (SPW) model—serve as fundamental baseline models for stochastic LoS channel characterization. By preserving the vectorial structure of Maxwell's equations and the dispersion relation, the framework naturally incorporates both propagating and evanescent modes. Our analysis of ergodic capacity and degrees of freedom (DoF) reveals that the key results of the complex SPW model can be reproduced by the simpler Gaussian model with limited variance. Furthermore, we provide examples using 2D continuous MIMO systems, illustrating how the model's Maxwell-consistent stochasticity explains observed increases
in channel capacity and DoF over the deterministic MIMO
capacity baseline. These idealized Maxwellian random field theoretic models offer a physically grounded reference point for understanding fundamental limits in stochastic LoS propagation environments.
\end{abstract}

\begin{IEEEkeywords}
 Stochastic Green's function, electromagnetic information, capacity, degree of freedom, multiple-input-multiple-output.

\end{IEEEkeywords}

\section{Introduction}

Multiple-input-multiple-output (MIMO) technology has long been recognized as a cornerstone of modern wireless communication systems for achieving high spectral efficiency, spatial diversity, and improved link reliability \cite{Telata1999MIMO_Capacity,marzetta2016fundamentals,9987524,9724113}. By exploiting the spatial domain of electromagnetic (EM) wave propagation, MIMO systems enable parallel data transmission across multiple antennas, thereby significantly enhancing capacity without additional bandwidth or power, particularly in line-of-sight (LoS) systems \cite{Dardari2021HMIMO,Gong2024HMIMO}. 
As wireless systems evolve toward higher carrier frequencies and larger array apertures, as in millimeter-wave and sub-terahertz communications \cite{lathi2019modern,10068425}, the physical behavior of EM propagation becomes increasingly intricate. In such regimes, the simplifying assumptions underlying conventional fading or geometric channel models, such as independent scattering and plane wave propagation, no longer remain valid \cite{9650519,10684477}. Instead, wave phenomena such as spatial correlation, polarization coupling, and near-field interactions become dominant, fundamentally altering the structure of the wireless channel \cite{10628002}. These effects challenge classical information theoretic formulations of capacity and motivate the need for physically consistent modeling approaches that accurately capture EM coupling between transceiver elements.

In the EM formulation of wireless propagation, the channel response between transmitting and receiving elements can be rigorously described by the dyadic Green’s function (DGF) of Maxwell’s equations \cite{tai1994dyadic,Chew_1990}, which captures polarization coupling, vectorial interactions, and both propagating and evanescent components. This representation reveals the spatial modal structure governing capacity, degree of freedom (DoF), and effective degree of freedom (eDoF) \cite{Mikki_book,clemmow1996the}, where eDoF represents the equivalent number of independent single-input-single-output (SISO) systems \cite{837052,4418491}. However, existing LoS DGF based studies are mostly deterministic \cite{1454881,c50041dcd40e486b9fdb9d09083b0024,10018012}. In practice, fluctuations in EM parameters (e.g., permittivity, environmental perturbations, fabrication tolerances) introduce randomness in the wavenumber \( k \), affecting both the Green’s function and MIMO performance \cite{8952896,9906802}, yet their influence remains insufficiently understood.

To address this gap, we propose a stochastic DGF (SDGF) framework that introduces physical randomness into the DGF while rigorously preserving Maxwellian structure. {The stochastic wavenumber is introduced as an effective representation of aggregated electromagnetic uncertainties, providing a canonical model for LoS-dominated scenarios with moderate randomness.} The key idea is to randomize the wavenumber \( k \), the fundamental parameter linking EM field propagation and spatial coherence. Two physically motivated stochastic models are investigated. The first is the classical Gaussian model, which assumes scalar wavenumber jitter and provides a convenient baseline for weakly random media\cite{ishimaru1997wave}. The second is our newly developed stochastic plane wave (SPW) model, derived from the plane wave spectrum expansion of the DGF\cite{felsen1994radiation,hansen1999plane_wave}. Unlike the Gaussian model, the SPW model randomizes the individual components of the wavevector \( \mathbf{k}=(k_1,k_2,k_3) \), thereby capturing both isotropic and anisotropic fluctuations and naturally incorporating propagation and evanescent spectral modes. {Such anisotropic fluctuations may arise from direction dependent electromagnetic uncertainties, including medium inhomogeneity, boundary roughness, and anisotropic propagation environments, which are commonly encountered in practical scenarios such as near-field large-aperture MIMO systems, LoS-dominated links with directional scattering (e.g., indoor or urban environments), and metasurface-assisted propagation. By representing these directional stochastic variations through componentwise wavevector randomization, the SPW model provides a physically grounded and richer description of stochastic EM propagation, especially in near-field and strongly coupled LoS scenarios. These statistical variations can, in practice, be inferred from measurable channel characteristics, such as spatial correlation functions or angular power spectra obtained via array-based channel sounding and beamforming techniques.} {Recent advances in electromagnetic information theory (EIT) have also provided new perspectives on stochastic EM channel modeling and DoF analysis\cite{10830535, 10417101}. However, this work constructs the channel directly from the DGF, thereby enabling a more physically grounded mapping through the communication link.}

The main contributions of this letter are summarized as follows. First, we propose the SPW model according to our established novel SDGF framework that introduces componentwise randomness into the EM wavenumber while strictly preserving Maxwellian consistency. Furthermore, when the statistical variances of the wavevector components are restricted, the SPW model can be reproduced by a simpler scalar Gaussian randomization, thereby achieving comparable stochastic behavior with significantly reduced computational complexity. Second, comprehensive Monte-Carlo simulations are performed to quantify how different fluctuation strengths and propagation distances affect modal richness and ergodic capacity. The results consistently demonstrate that introducing controlled wavenumber randomness leads to a substantial improvement in ergodic capacity and spatial DoF compared to the deterministic LoS case. {These results show how stochastic EM parameters and the propagating/evanescent modal partition jointly affect modal diversity, DoF, and ergodic capacity.}


\section{The SPW Framework for Capacity in Continious MIMO Systems}

In the frequency domain, time-harmonic EM fields are represented as complex-valued quantities with $\exp(-\im\omega t)$ dependence \cite{Van_Bladel_2007}. For a spatial Fourier mode \( \exp(\im\mathbf{k}\cdot\mathbf{x} - i\omega t) \), the dispersion relation in a homogeneous, isotropic medium is given by $\|\mathbf{k}\|=k=k_0 \epsilon_{\rm eff}$, where $k_0=\omega/c$ is the vacuum wavenumber, \( c \) is the speed of light, and $\epsilon_{\rm eff}$ is the effective permittivity \cite{landau1984electrodynamics,schwinger1998classical}. 

The MIMO channel is characterized by DGF, which captures vectorial field coupling and polarization effects in LoS propagation. To incorporate electromagnetic uncertainty, we introduce stochasticity into the wavenumber, rendering the Green’s function random. For tractability, we consider a homogeneous, isotropic, non-magnetic, and lossless medium, where $\epsilon_{\rm eff}$ is modeled as a random variable rather than a spatial random field, while retaining the full vectorial DGF formulation.
For a medium characterized by \( \mu = \mu_0 \), \( \epsilon = \epsilon_0 \epsilon_{\rm eff} \), and wavenumber \( k \), the solution can also be written as \cite{tai1994dyadic}
\begin{equation}\label{E_rad DGF}
\begin{aligned}
\overline{\mathbf{G}}(\mathbf{x} ,\mathbf{x} ';k ) &= \gamma_{\omega} g_1(R)\left(\overline{\mathbf{I}} -\hat{R}\otimes\hat{R}\right)  \\
&\quad -\gamma_{\omega} \left[g_2(R) -g_3(R)\right]\left(\overline{\mathbf{I}} -3\hat{R}\otimes\hat{R}\right),
\end{aligned}
\end{equation}
where
\( g_n(\mathbf{x}, \mathbf{x}'; k) := ke^{\im k R}/(i k R)^n, \ \ n \in \{1,2,3\} \),
\(\gamma_{\omega} := -\mu _{0} \omega /4\pi\), \( \mathbf{R} := \mathbf{x} - \mathbf{x}' \), \( R := \norm{\mathbf{R}} \), and \( \hat{R} := \mathbf{R}/R \).

Within the proposed SDGF framework (to be detailed subsequently), both propagating and evanescent field components must adhere to Maxwell's equations, thereby satisfying the classical dispersion relation \cite{Chew_1990,Wald2022electromagnetism}: $k_1^2 + k_2^2 + k_3^2 = k^2 = k_0^2 n^2 = k_0^2 \epsilon_{\mathrm{eff}}$, where $\mathbf{k} = \sum_{i=1}^3 \hat{\mathbf{x}}_i k_i$ is the wave vector, $n = \sqrt{\epsilon_{\mathrm{eff}}}$ is the refractive index \cite{ishimaru1997wave}, and the corresponding wavelength is $\lambda = 2\pi/k$. The introduction of stochasticity in $k$ consequently renders $\lambda$ a random variable. Furthermore, to ensure physical realizability, the probability distribution of the random wavenumber must be constrained such that $\mathbb{P}(k \leq 0) \simeq 0$.
Guided by these constraints on the randomization of $k$, we formulate two specific SDGF models: the Gaussian model and the SPW model. The Gaussian model provides a natural representation of a random effective refractive index, positing $k = k_0 + \Delta k$, with $\Delta k \sim \mathcal{N}(0, \sigma^2)$. Its principal advantage is its tractability to full analytical treatment via probability theory, a comprehensive analysis of which will be presented elsewhere. While this model is typically considered accurate for weakly random media, it has also found application in select scenarios with strong fluctuations \cite{ishimaru1997wave}. 

A naive stochastic model might simply randomize the global wavenumber $k$ as a scalar, $k = k_0 + \Delta k$. However, such an approach is electromagnetically inconsistent because it severs the critical link between the wavevector's direction and its magnitude. By randomizing the wavevector components $k_i$, our SPW model preserves this physical constraint, which allows directional stochasticity and reactive near-field effects to be represented consistently. As a result, stochastic uncertainties are mapped into a physically consistent plane wave spectrum containing both propagating and evanescent components.
{It should be emphasized that the scalar Gaussian model is introduced only as a reduced surrogate for weak stochastic perturbations, and not as a universal representation of wave propagation in arbitrary random media.} To fully capture the EM behavior of stochastic LoS propagation, we introduce a novel SPW model. This model is derived from the well-known plane-wave spectrum expansion of the DGF \cite{Kerns1976,clemmow1996the}. Inded, it is now generally recognized that the decomposition of the EM field into propagating and non-propagating parts is fundamental for understanding not only near-field phenomena \cite{mikki_nf2, hansen1999plane_wave} but also mutual coupling \cite{rigorous_MC, mikki_energy, Kerns1976, Kerns_1976}, interactions, and information transmission \cite{Mikki_book}.

In contrast to the conventional Gaussian model, which randomizes only the scalar wavenumber $k$, the proposed SPW model treats the wavevector $\mathbf{k} = [k_1, k_2, k_3]^\top$ itself as a random variable. In this formulation, each Cartesian component follows a Gaussian distribution, permitting either isotropic or anisotropic fluctuations. This approach naturally incorporates both propagating and evanescent spectral modes, thereby ensuring a Maxwell-consistent representation of near-field and mutual coupling effects. By randomizing the directional components of the wavevector, the SPW model provides a physically grounded and computationally efficient framework capable of capturing the spatial richness and stochastic fluctuations of EM fields across all propagation regimes. {In this sense, the SDGF framework constitutes a physically consistent stochastic extension of the DGF, rather than a purely phenomenological perturbation.}
To implement the SPW model in an electromagnetically consistent manner, the dispersion relation $\norm{\vb{k}} = \omega/c$ must be strictly enforced. We achieve this by generating a random wavevector $\vb{k}$ form consistent with the propagating/evanescent mode decomposition:
\begin{equation}\label{eq:RV_model_k_letter}
    k = \begin{cases}
        \sqrt{k_1^2 + k_2^2 - k_3^2}, & k_1^2 + k_2^2 > k_3^2 \quad \text{(Evanescent)}, \\
        \sqrt{k_1^2 + k_2^2 + k_3^2}, & k_3^2 \ge k_1^2 + k_2^2 \quad \text{(Propagating)}.
    \end{cases}
\end{equation}
where \( k_i \sim \mathcal{N}(\mu_i, \sigma_i^2) \) for \( i \in \{1, 2, 3\} \). Here, $k_3 = \sqrt{k_0^2 n^2 - k_1^2 - k_2^2}$ is real in the visible (propagating) region and purely imaginary in the invisible (evanescent) region \cite{ishimaru1997wave,Chew_1990}. {It is important to mention that the Gaussian model can be interpreted as a first order approximation of the SPW model in the small fluctuation regime. This corresponds to the case where the relative variation $\sigma_i/\mu_i$ remains sufficiently small, so that the nonlinear mapping from the wavevector components to the effective wavenumber can be accurately approximated by its first order Taylor expansion, and branch switching between propagating and evanescent modes remains infrequent. In this sense, the Gaussian approximation is quantitatively reliable for $\sigma_i/\mu_i \lesssim 0.2$ and remains qualitatively accurate up to approximately $\sigma_i/\mu_i \lesssim 0.3$.} This formulation yields a physically consistent SDGF that rigorously incorporates both spectral regimes while remaining computationally tractable for the capacity evaluation that follows.

\begin{figure}[!h]
    \centering
    \includegraphics[width=0.35\textwidth ]{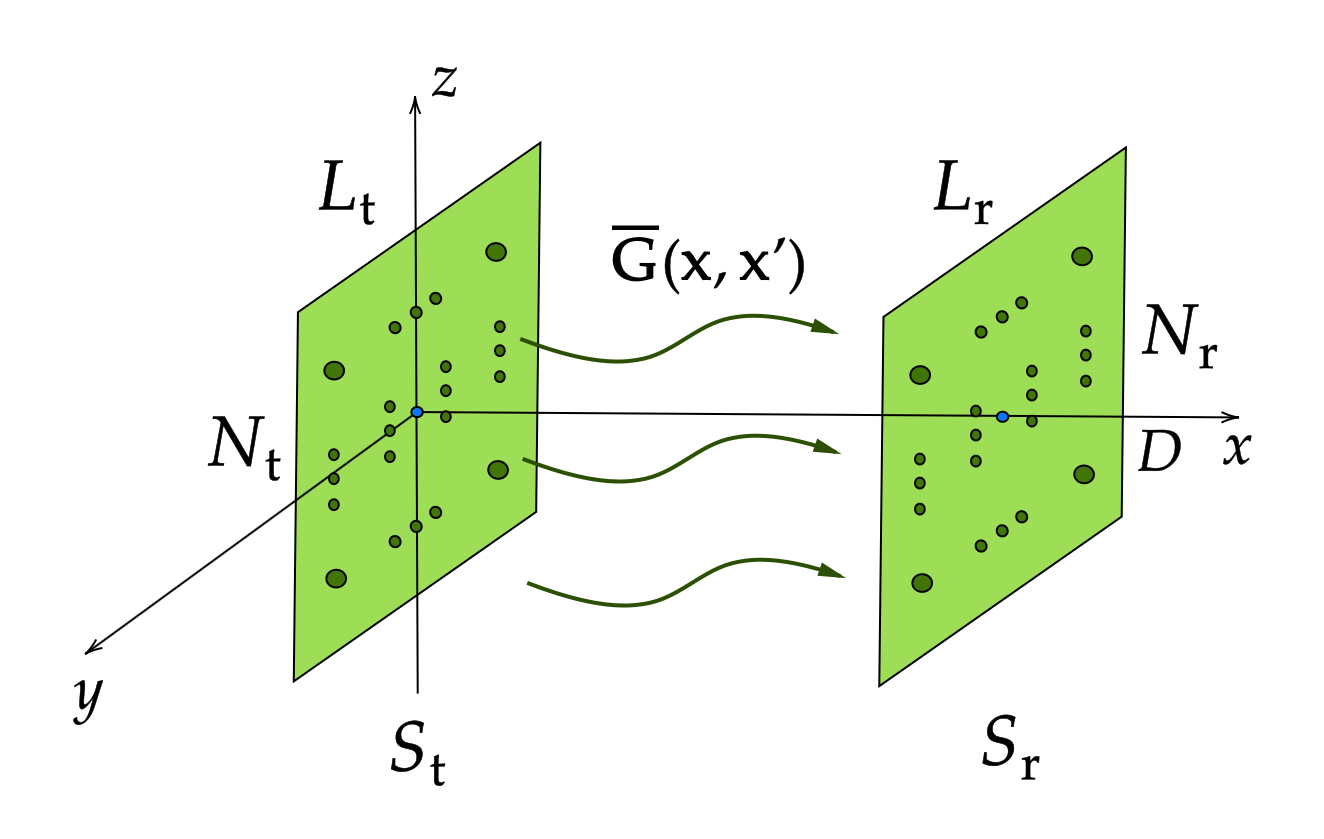}
    \caption{Our defined continuous MIMO system consists of $N_\mathrm{t}\cross N_\mathrm{r}$ uniformly distributed point transmitter $S_\mathrm{t}$ and receiver $S_\mathrm{r}$ on two identical square planes. $L$ is the side length of the planes, and $D$ is the distance between the two planes.}
    \label{fig:cMIMO_system}
\end{figure}

For a MIMO configuration with $N_\mathrm{t}$ transmitting and $N_\mathrm{r}$ receiving elements (Fig.~\ref{fig:cMIMO_system}), the overall EM channel is described by a $3N_\mathrm{r} \times 3N_\mathrm{t}$ matrix $\widebar{\widebar{H}}$. This matrix is constructed by stacking the DGF tensors for each transmitter-receiver pair. Specifically, each submatrix $\mathbf{H}_{pq}$ (with $p,q \in \{x,y,z\}$) within $\widebar{\widebar{H}}$ contains the DGF values governing the coupling between all $N_\mathrm{r}$ receive and $N_\mathrm{t}$ transmit elements for $p$-polarized reception and $q$-polarized excitation. This structure fully encapsulates the vectorial propagation characteristics of the channel.
The complete channel matrix $\widebar{\widebar{H}}$ can be randomized using either the Gaussian or the Stochastic SPW model introduced previously. {For numerical implementation, the stochastic channel matrix is assembled on a Monte-Carlo basis rather than through a truncated summation of plane wave terms. For the $m$th realization, an effective stochastic wavenumber $k^{m}$ is first generated according to the selected stochastic model, i.e., the scalar Gaussian model or the SPW model. The corresponding channel matrix is then constructed in a polarization-resolved form as
$
\big[\mathbf{H}_{pq}^{(m)}\big]_{j,i}
=
\mathbf{e}_p^\top\cdot
\overline{\mathbf{G}}\!\left(\mathbf{x}_j,\mathbf{x}_i;k^{(m)}\right)
\cdot\mathbf{e}_q,$
where $i \in \{1,\dots,N_\mathrm{t}\}$ and $j \in \{1,\dots,N_\mathrm{r}\}$ index the transmitting and receiving elements, respectively, and $\mathbf{e}_p$ and $\mathbf{e}_q$ denote the unit vectors along the $p$ and $q$ directions, respectively.} Once the stochastic channel matrix is obtained, performance metrics like capacity and DoF can be derived directly from its eigenvalue spectrum. {It is noted that the channel matrix is first constructed directly from the DGF and therefore carries the physical dimensions inherited from the electromagnetic Green's function. For capacity and DoF evaluation, it is normalized by its Frobenius norm, yielding a dimensionless effective channel with fixed total channel energy. Thus, the reported capacity variations are not due to absolute power scaling or loss of energy conservation, but reflect changes in the spatial structure and eigenvalue distribution of the channel.}
For an AWGN MIMO system, the instantaneous channel capacity per unit bandwidth is given by $C = \log_2 \det\!\left[ \mathbf{I}_{N_\mathrm{r}} + \rho\, \widebar{\widebar{H}}\, \widebar{\widebar{H}}^\dagger \right] \ \mathrm{(bit/s/Hz)}$,
where $\rho$ is the signal-to-noise ratio (SNR) \cite{Telata1999MIMO_Capacity,10012689}. Under the assumptions of statistically independent transmit symbols and uncorrelated noise, the channel matrix $\widebar{\widebar{H}}$ directly embodies the EM coupling between the transmitting and receiving surfaces via the stochastic Green's tensor. This expression thus characterizes the fundamental limit of information transmission under the proposed stochastic propagation model.
Building upon the stochastic EM channel matrix $\widebar{\widebar{H}}$ formulated above, we next assess the system's spatial multiplexing capability by quantifying its intrinsic modal richness. This is characterized through two complementary measures: the number of DoF, which counts the statistically independent spatial subchannels, and the eDoF, a continuous measure that weights each eigenmode by its energy contribution, thus reflecting practical multiplexing efficiency.
Within this framework, the DoF is computed by thresholding the normalized singular values of $\widebar{\widebar{H}}$: $\mathrm{DoF} = \big| \{\, \varsigma_i : \varsigma_i / \varsigma_{\max}(\widebar{\widebar{H}}) \ge \zeta \,\} \big|$, where $\varsigma_i$ are the singular values, $\zeta$ is a predefined threshold (here, $\zeta = 0.05$), and $|\cdot|$ denotes set cardinality. The eDoF is defined as $\mathrm{eDoF} = \left[ \mathrm{Tr}\!\left( \widebar{\widebar{H}} \, \widebar{\widebar{H}}^{\dagger} \right) / \left\lVert \widebar{\widebar{H}} \, \widebar{\widebar{H}}^{\dagger} \right\rVert_F \right]^2$.
Together, these metrics provide deeper physical insight into how EM coupling and wavenumber randomness jointly influence the achievable spatial capacity of the MIMO system. In particular, increased randomness in the Green's function (e.g., larger variances $\sigma_i$ in the SPW model) tends to enrich the singular value spectrum, increasing the number of dominant modes and thereby enhancing both the channel capacity and the effective spatial diversity.

\section{Simulation Results and Analysis}

\begin{figure}[htbp]
  \centering
  \begin{subfigure}[b]{0.25\textwidth}
    \centering
    \includegraphics[width=\linewidth]{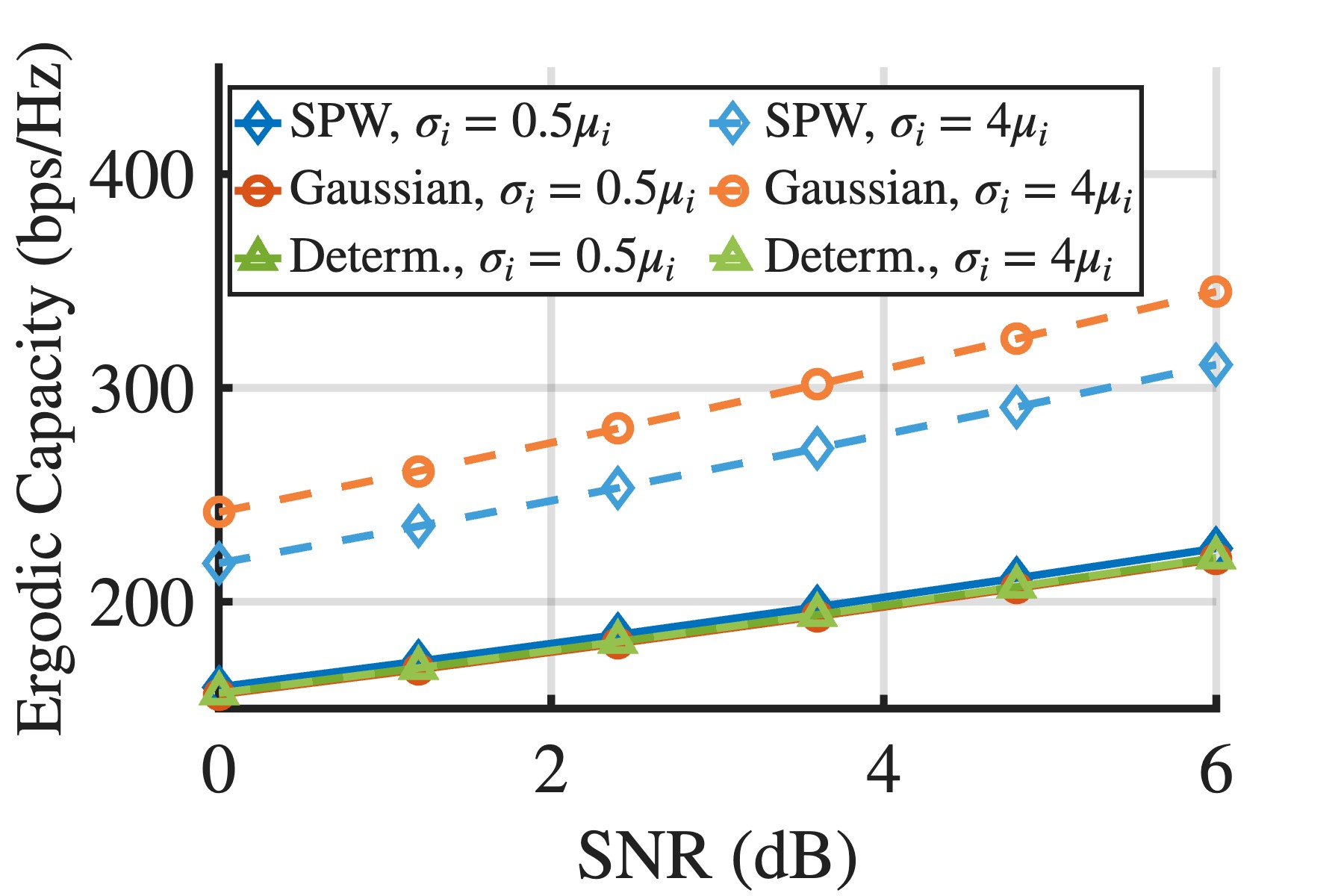}
    \caption{$d =0.2\lambda$}
    \label{fig:sigma_sub1}
  \end{subfigure}
  \hspace{-0.05\linewidth}             
  \begin{subfigure}[b]{0.25\textwidth}
    \centering
    \includegraphics[width=\linewidth]{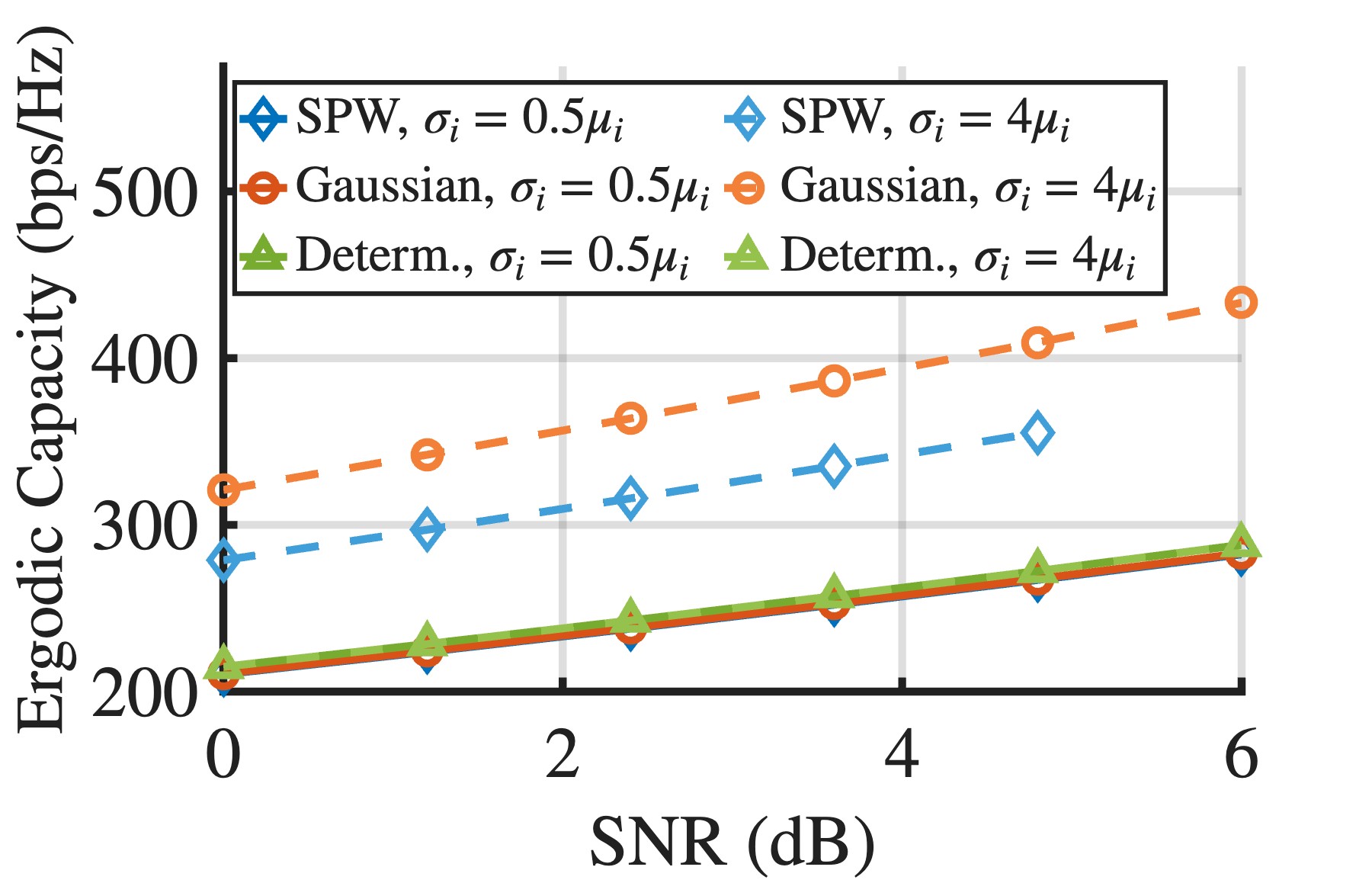}
    \caption{$d =0.1\lambda$}
    \label{fig:sigma_sub2}
  \end{subfigure}
  \caption{The comparison of the ergodic capacity of the cMIMO system of SPW, Gaussian and deterministic model with $\sigma_i {/} \mu_i = 0.5$ and $\sigma_i {/} \mu_i = 4$, where the size of the plane transceivers is fixed at $L_\mathrm{t} = L_\mathrm{r}=5\lambda$, the distance between them is $D = 30\lambda$, the inter-element spacing is $d_\mathrm{t,x}=0.2\lambda$ for Fig. \ref{fig:sigma_sub1} and $d_\mathrm{t,x}=0.1\lambda$ for Fig. \ref{fig:sigma_sub2}. The range of SNR varies from 0 to 6dB with operating frequency at 6\,GHz.}
  \label{fig:Capacity_30lambda_different_sigma}
\end{figure}

\begin{figure}[htbp]
  \centering
  \begin{subfigure}[b]{0.25\textwidth}
    \centering
    \includegraphics[width=\linewidth]{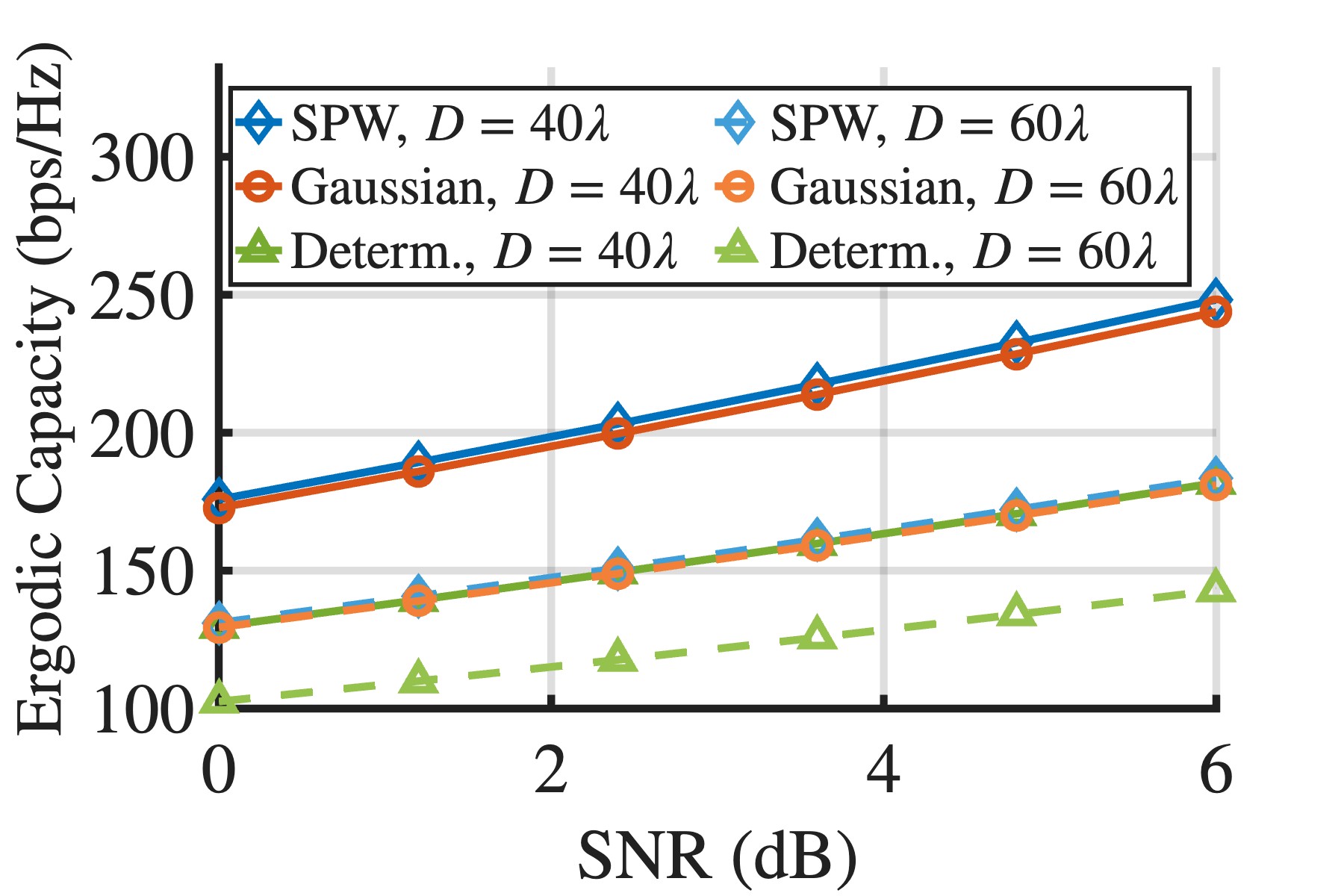}
    \caption{$d =0.2\lambda$}
    \label{fig:Dsub1}
  \end{subfigure}
  \hspace{-0.05\linewidth}             
  \begin{subfigure}[b]{0.25\textwidth}
    \centering
    \includegraphics[width=\linewidth]{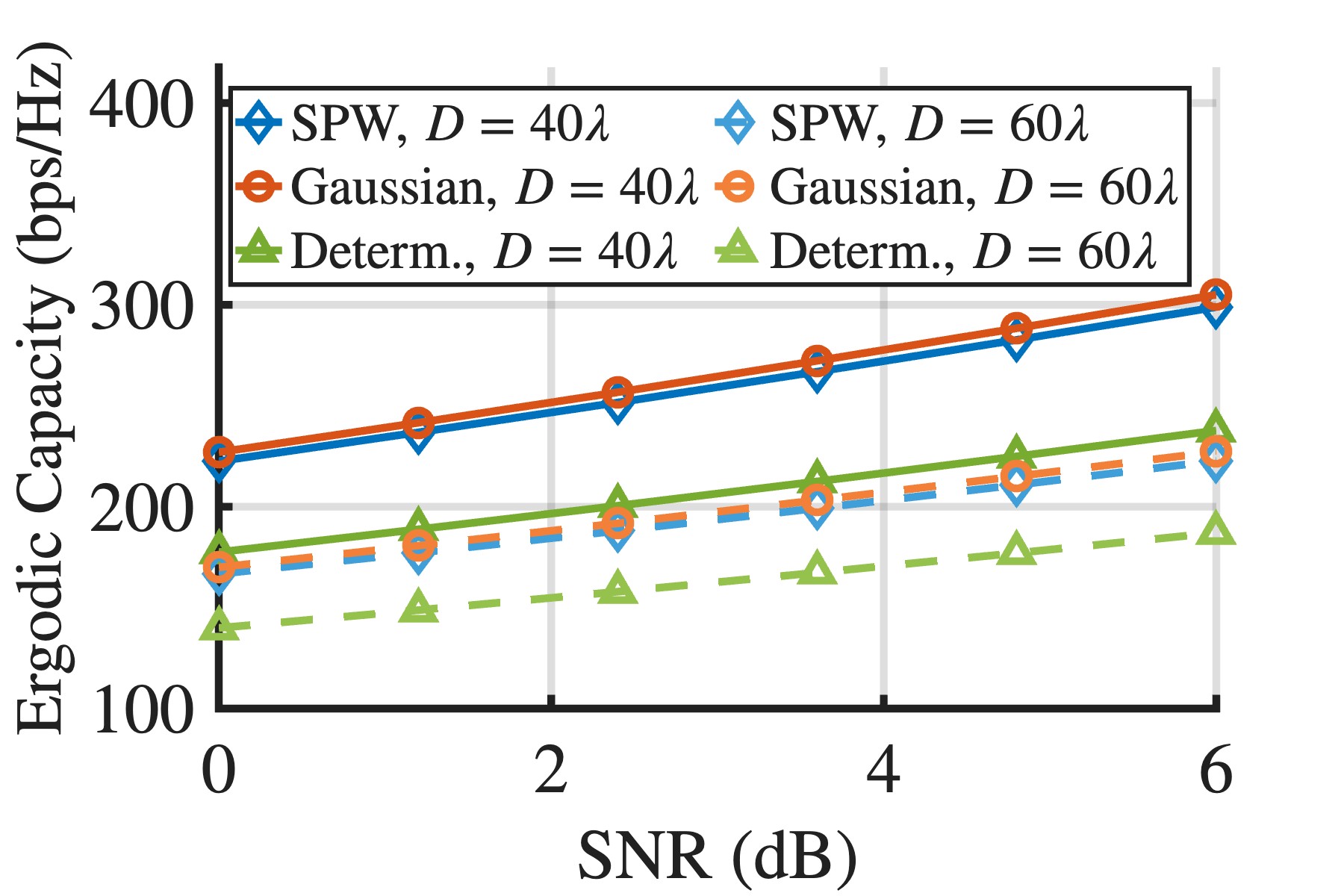}
    \caption{$d =0.1\lambda$}
    \label{fig:Dsub2}
  \end{subfigure}
  \caption{Same as Fig. \ref{fig:Capacity_30lambda_different_sigma}, with $\sigma_i {/} \mu_i = 4$ and the distance between them changes between NF $D = 40\lambda$ and FF $D = 60\lambda$. The inter-element spacing is $d_\mathrm{t,x}=0.2\lambda$ for Fig. \ref{fig:Dsub1} and $d_\mathrm{t,x}=0.1\lambda$ for Fig. \ref{fig:Dsub2}.}
  \label{fig:Capacity_sigma4mu}
\end{figure}

This section presents Monte Carlo simulation results for the ergodic capacity under both the Gaussian and SPW models, benchmarked against a deterministic LoS baseline. The objective is to validate the overall scaling behavior and relative performance trends discussed previously. {Here, the Gaussian model is constructed by matching the empirical mean and variance of the SPW-generated wavenumber.}
Figs.~\ref{fig:Capacity_30lambda_different_sigma} and \ref{fig:Capacity_sigma4mu} illustrate the ergodic capacity for a continuous MIMO (cMIMO) system with aperture sizes \( L_\mathrm{t} = L_\mathrm{r} = 5\lambda \) at 6 GHz. The results demonstrate consistent and physically interpretable trends across variations in element spacing, transceiver distance, and fluctuation strength. A key observation is the noticeable growth in ergodic capacity as the inter-element spacing \( d_\mathrm{t,x} \) decreases from \( 0.2\lambda \) to \( 0.1\lambda \) for all models. This is attributed to enhanced spatial correlation, which increases the number of effective transmission modes. Furthermore, increasing the stochastic fluctuation strength from \( \sigma_i=0.5\mu_i \) to \( \sigma_i=4\mu_i \) yields additional capacity gains for both stochastic models, underscoring the constructive role of moderate EM randomness in improving the effective rank of the channel matrix. The performance gap between the SPW and Gaussian models widens at higher fluctuation levels, as the SPW model's perturbation of the entire wavevector provides a more complete representation of EM stochasticity compared to the Gaussian model's randomization of only the scalar wavenumber.
In Fig.~\ref{fig:Capacity_sigma4mu}, with the fluctuation strength fixed at \( \sigma_i=4\mu_i \), the capacity decreases as the transceiver distance increases from the near-field (NF, \( D=40\lambda \)) to the far-field (FF, \( D=60\lambda \)). {Since the channel matrix is normalized before capacity evaluation, the observed capacity reduction is structural rather than due to path loss, and mainly stems from the progressive attenuation of the evanescent spectrum.} These modes, corresponding to the spectral region where \( k_\rho = \sqrt{k_1^2 + k_2^2} > k \), experience an exponential decay \( \exp(-D \sqrt{k_\rho^2 - k^2}) \) with distance \( D \). {While individual evanescent components decay rapidly with distance, they collectively contribute to the high spatial frequency content of the field, and their attenuation effectively acts as a spectral filtering mechanism.} Consequently, the associated reactive near-field components (characterized by \( 1/R^2 \) and \( 1/R^3 \) dependence in the DGF) are progressively suppressed, and the channel becomes dominated by a smaller set of propagating modes (\( k_\rho \leq k \)) \cite{10005192}. {This progressive reduction of the effective spatial bandwidth leads to a more concentrated singular value distribution and a decrease in effective degrees of freedom,} which directly results in the observed reduction in ergodic capacity.

\begin{figure}[htbp]
  \centering
  \begin{subfigure}[b]{0.25\textwidth}
    \centering
    \includegraphics[width=\linewidth]{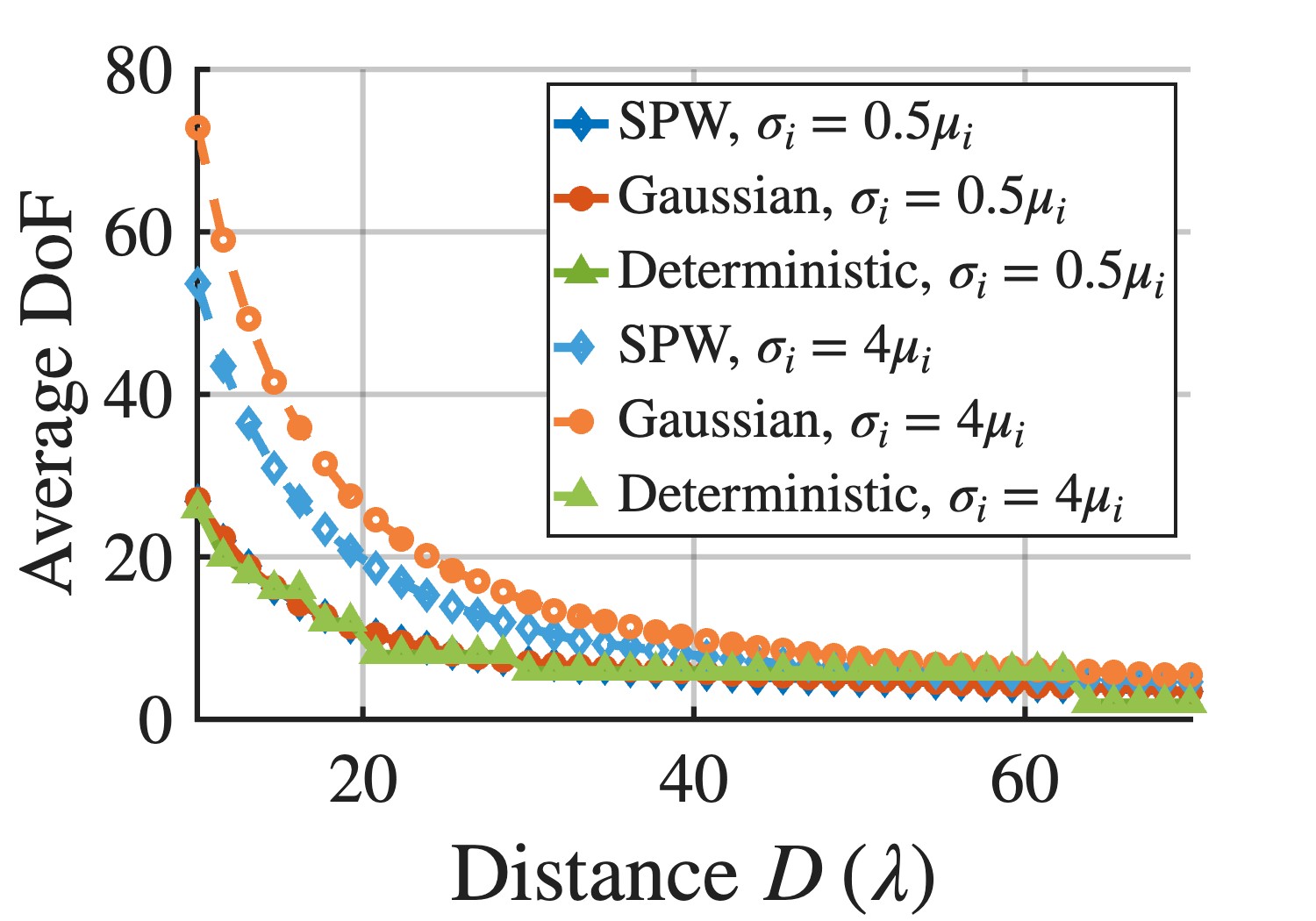}
    \caption{DoF}
    \label{fig:DoF}
  \end{subfigure}
  \hspace{-0.05\linewidth}             
  \begin{subfigure}[b]{0.25\textwidth}
    \centering
    \includegraphics[width=\linewidth]{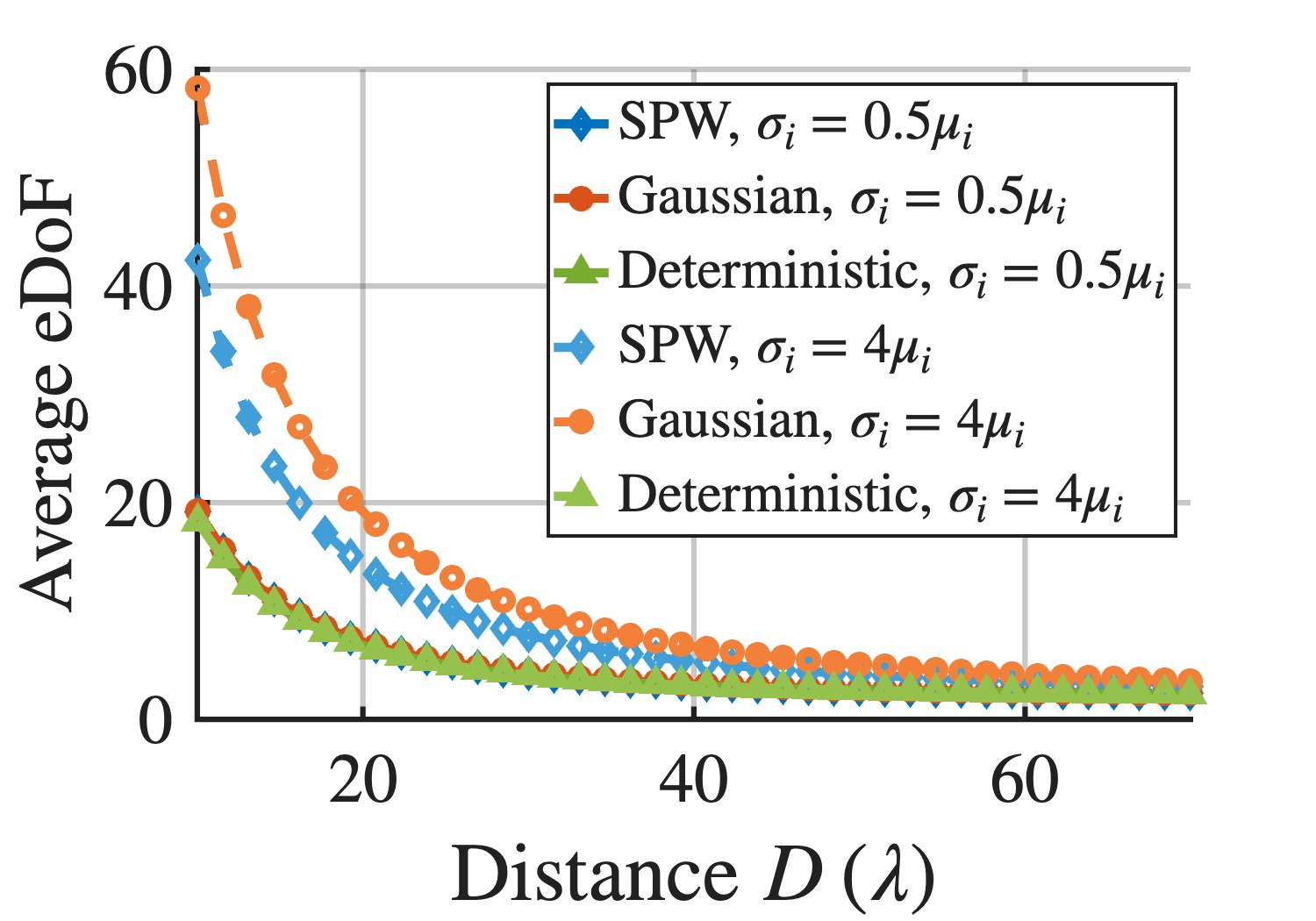}
    \caption{eDoF}
    \label{fig:eDoF}
  \end{subfigure}
  \caption{The comparison of DoF and eDoF for the stochastic Green's Function and deterministic Green's Function based MIMO system with fixed aperture size $L_\mathrm{t} = L_\mathrm{r} = 5\lambda$, inter-element spacings $d_\mathrm{t,r} =0.2\lambda$ and $\sigma_i {/} \mu_i \in \{ 0.5, 4\}$ for SPW model, Tx-Rx distance $D$ changing from NF to FF, and operating frequency 6\,GHz.}
  \label{fig:DoF_eDoF}
\end{figure}

Fig. \ref{fig:DoF_eDoF} illustrates the variation of the average DoF and eDoF as functions of the transceiver separation distance $D$ for both stochastic and deterministic channel models, Fig. \ref{fig:DoF} is for DoF while Fig. \ref{fig:eDoF} describes eDoF. In all cases, the DoF decreases monotonically with the increasing of $D$. This behavior stems from the gradual contraction of the channel’s spatial spectrum as the separation grows, at shorter distance, the EM field supports a rich mixture of propagating and evanescent spectral components that form a dense set of spatial modes. When $D$ increases, higher spatial frequency components are progressively filtered out, leaving only a limited subset of long range
propagating modes. Consequently, the singular value distribution of $\overline{\overline{H}}$ becomes increasingly concentrated, and the effective modal dimensionality shrinks. This spectral truncation with distance, rather than a simple weakening of coupling strength, fundamentally accounts for the observed monotonic reduction of DoF. For short range configurations, the SPW model and the Gaussian model yield significantly larger DoF and eDoF, followed by the deterministic model. This indicates that the componentwise randomization of the wavevector in our SDGF models preserves a richer set of spatial field variations. As the fluctuation strength increases from $\sigma_i=0.5\mu_i$ to $\sigma_i=4\mu_i$, both stochastic models exhibit an initial DoF and eDoF enhancement, after which the curves converge toward a common asymptotic limit dominated by propagating modes. 

Beyond this general decay trend, the DoF and eDoF curves for all models converge to similar values as $D$ increases into the far-field (FF) regime. This implies that stochastic field fluctuations primarily influence the near-field (NF) and intermediate transition regions, with a diminishing impact in the FF. Furthermore, the distinction between DoF and eDoF vanishes with increasing distance. Notably, in the FF, both metrics for all models approach a value of 2, which aligns with the fundamental limit for a planar cMIMO system dictated by the two polarization directions of a plane wave \cite{Dardari2020HMIMO}. Additionally, the eDoF consistently lies slightly below the DoF across all distances, indicating that only a subset of the available spatial modes contribute significantly to the total received power. 
{The proposed SDGF framework is complementary to conventional statistical fading models. Classical Rayleigh/Rician models describe environment-induced randomness caused by multipath scattering, whereas our SDGF model introduces EM-aware randomness directly through the Green's function and the wavenumber perturbations. A systematic comparison with conventional fading models and a detailed study of hybrid formulations combining the two kinds of randomness will be pursued in future work.} 

\section{Conclusion}
This letter has shown that stochastic LoS MIMO propagation can be modeled through minimal Maxwellian random field perturbations of the dyadic Green's function. The Gaussian and SPW constructions represent two complementary levels of randomness: the former gives a tractable scalar baseline, while the latter retains componentwise wavevector fluctuations and the associated propagating/evanescent modal structure. The 2D continuous MIMO examples indicate that these models mainly affect the singular value distribution of the channel, leading to changes in normalized capacity and DoF. In this sense, the proposed models provide a compact reference for studying how Maxwell-consistent stochasticity reshapes LoS MIMO channel structure.


\begin{thebibliography}{10}
\providecommand{\url}[1]{#1}
\csname url@samestyle\endcsname
\providecommand{\newblock}{\relax}
\providecommand{\bibinfo}[2]{#2}
\providecommand{\BIBentrySTDinterwordspacing}{\spaceskip=0pt\relax}
\providecommand{\BIBentryALTinterwordstretchfactor}{4}
\providecommand{\BIBentryALTinterwordspacing}{\spaceskip=\fontdimen2\font plus
\BIBentryALTinterwordstretchfactor\fontdimen3\font minus \fontdimen4\font\relax}
\providecommand{\BIBforeignlanguage}[2]{{%
\expandafter\ifx\csname l@#1\endcsname\relax
\typeout{** WARNING: IEEEtran.bst: No hyphenation pattern has been}%
\typeout{** loaded for the language `#1'. Using the pattern for}%
\typeout{** the default language instead.}%
\else
\language=\csname l@#1\endcsname
\fi
#2}}
\providecommand{\BIBdecl}{\relax}
\BIBdecl

\bibitem{Telata1999MIMO_Capacity}
E.~Telatar, ``Capacity of multi-antenna {Gaussian} channels,'' \emph{European Transactions on Telecommunications}, vol.~10, no.~6, pp. 585--595, 1999.

\bibitem{marzetta2016fundamentals}
T.~L. Marzetta, E.~G. Larsson, H.~Yang, and H.~Q. Ngo, \emph{Fundamentals of Massive {MIMO}}.\hskip 1em plus 0.5em minus 0.4em\relax Cambridge University Press, 2016.

\bibitem{9987524}
Y.~Jiang and F.~Gao, ``Electromagnetic channel model for near field mimo systems in the half space,'' \emph{IEEE Communications Letters}, vol.~27, no.~2, pp. 706--710, 2023.

\bibitem{9724113}
A.~Pizzo, L.~Sanguinetti, and T.~L. Marzetta, ``Fourier plane-wave series expansion for holographic mimo communications,'' \emph{IEEE Transactions on Wireless Communications}, vol.~21, no.~9, pp. 6890--6905, 2022.

\bibitem{Dardari2021HMIMO}
D.~Dardari and N.~Decarli, ``Holographic communication using intelligent surfaces,'' \emph{IEEE Communications Magazine}, vol.~59, no.~6, pp. 35--41, 2021.

\bibitem{Gong2024HMIMO}
T.~Gong, P.~Gavriilidis, R.~Ji, C.~Huang, G.~C. Alexandropoulos, L.~Wei, Z.~Zhang, M.~Debbah, H.~V. Poor, and C.~Yuen, ``Holographic {MIMO} communications: Theoretical foundations, enabling technologies, and future directions,'' \emph{IEEE Communications Surveys \& Tutorials}, vol.~26, no.~1, pp. 196--257, 2024.

\bibitem{lathi2019modern}
B.~P. Lathi and Z.~Ding, \emph{Modern digital and analog communication systems}.\hskip 1em plus 0.5em minus 0.4em\relax New York: Oxford University Press, 2019.

\bibitem{10068425}
Z.~Han, S.~Shen, Y.~Zhang, S.~Tang, C.-Y. Chiu, and R.~Murch, ``Using loaded n-port structures to achieve the continuous-space electromagnetic channel capacity bound,'' \emph{IEEE Transactions on Wireless Communications}, vol.~22, no.~11, pp. 7592--7605, 2023.

\bibitem{9650519}
S.~S.~A. Yuan, Z.~He, X.~Chen, C.~Huang, and W.~E.~I. Sha, ``Electromagnetic effective degree of freedom of an mimo system in free space,'' \emph{IEEE Antennas and Wireless Propagation Letters}, vol.~21, no.~3, pp. 446--450, 2022.

\bibitem{10684477}
Z.~Wan, J.~Zhu, and L.~Dai, ``Near-field channel modeling for electromagnetic information theory,'' \emph{IEEE Transactions on Wireless Communications}, vol.~23, no.~12, pp. 18\,004--18\,018, 2024.

\bibitem{10628002}
R.~Ji, C.~Huang, X.~Chen, W.~E.~I. Sha, L.~Dai, J.~He, Z.~Zhang, C.~Yuen, and M.~Debbah, ``Electromagnetic hybrid beamforming for holographic mimo communications,'' \emph{IEEE Transactions on Wireless Communications}, vol.~23, no.~11, pp. 15\,973--15\,986, 2024.

\bibitem{tai1994dyadic}
C.-T. Tai, \emph{Dyadic green functions in electromagnetic theory}.\hskip 1em plus 0.5em minus 0.4em\relax Piscataway, NJ: IEEE Press, 1994.

\bibitem{Chew_1990}
W.~C. Chew, \emph{{Waves and fields in inhomogenous media}}.\hskip 1em plus 0.5em minus 0.4em\relax Wiley-IEEE, 1999.

\bibitem{Mikki_book}
S.~Mikki and Y.~Antar, \emph{New Foundations for Applied Electromagnetics: The Spatial Structure of Fields}.\hskip 1em plus 0.5em minus 0.4em\relax London: Artech House, 2016.

\bibitem{clemmow1996the}
P.~C. Clemmow, \emph{The Plane Wave Spectrum Representation of Electromagnetic Fields}.\hskip 1em plus 0.5em minus 0.4em\relax New York Oxford: Institute of Electrical and Electronics Engineers Oxford University Press, 1996.

\bibitem{837052}
D.-S. Shiu, G.~Foschini, M.~Gans, and J.~Kahn, ``Fading correlation and its effect on the capacity of multielement antenna systems,'' \emph{IEEE Transactions on Communications}, vol.~48, no.~3, pp. 502--513, 2000.

\bibitem{4418491}
T.~Muharemovic, A.~Sabharwal, and B.~Aazhang, ``Antenna packing in low-power systems: Communication limits and array design,'' \emph{IEEE Transactions on Information Theory}, vol.~54, no.~1, pp. 429--440, 2008.

\bibitem{1454881}
A.~Ishimaru, ``Theory and application of wave propagation and scattering in random media,'' \emph{Proceedings of the IEEE}, vol.~65, no.~7, pp. 1030--1061, 1977.

\bibitem{c50041dcd40e486b9fdb9d09083b0024}
H.~P{\'e}cseli, ``\BIBforeignlanguage{English}{Electromagnetic wave propagation in random media},'' \emph{\BIBforeignlanguage{English}{Modern Physics Letters A}}, vol. 105, no.~9, pp. 468--471, 1984.

\bibitem{10018012}
S.~Lin, S.~Luo, S.~Ma, J.~Feng, Y.~Shao, Z.~B. Drikas, B.~D. Addissie, S.~M. Anlage, T.~Antonsen, and Z.~Peng, ``Predicting statistical wave physics in complex enclosures: A stochastic dyadic green's function approach,'' \emph{IEEE Transactions on Electromagnetic Compatibility}, vol.~65, no.~2, pp. 436--453, 2023.

\bibitem{8952896}
S.~Lin, Z.~Peng, and T.~M. Antonsen, ``A stochastic green's function for solution of wave propagation in wave-chaotic environments,'' \emph{IEEE Transactions on Antennas and Propagation}, vol.~68, no.~5, pp. 3919--3933, 2020.

\bibitem{9906802}
L.~Sanguinetti, A.~A. D'Amico, and M.~Debbah, ``Wavenumber-division multiplexing in line-of-sight holographic mimo communications,'' \emph{IEEE Transactions on Wireless Communications}, vol.~22, no.~4, pp. 2186--2201, 2023.

\bibitem{ishimaru1997wave}
A.~Ishimaru, \emph{Wave propagation and scattering in random media}.\hskip 1em plus 0.5em minus 0.4em\relax New York Oxford New York: IEEE Press Oxford University Press, 1997.

\bibitem{felsen1994radiation}
L.~Felsen, \emph{Radiation and scattering of waves}.\hskip 1em plus 0.5em minus 0.4em\relax Piscataway, NJ: IEEE Press, 1994.

\bibitem{hansen1999plane_wave}
T.~Hansen and A.~D. Yaghjian, \emph{Plane-Wave Theory of Time-Domain Fields: Near-Field Scanning applications}.\hskip 1em plus 0.5em minus 0.4em\relax New York: IEEE Press, 1999.

\bibitem{10830535}
J.~Zhu, Z.~Wan, L.~Dai, and T.~Jun~Cui, ``Electromagnetic information theory-based statistical channel model for improved channel estimation,'' \emph{IEEE Transactions on Information Theory}, vol.~71, no.~3, pp. 1777--1793, 2025.

\bibitem{10417101}
J.~Zhu, Z.~Wan, L.~Dai, M.~Debbah, and H.~V. Poor, ``Electromagnetic information theory: Fundamentals, modeling, applications, and open problems,'' \emph{IEEE Wireless Communications}, vol.~31, no.~3, pp. 156--162, 2024.

\bibitem{Van_Bladel_2007}
J.~Bladel, \emph{{Electromagnetic fields}}.\hskip 1em plus 0.5em minus 0.4em\relax Wiley-Intersience, IEEE, 2007.

\bibitem{landau1984electrodynamics}
L.~D. Landau, E.~Lifshitz, and L.~P. Pitaevskii, \emph{Electrodynamics of continuous media}.\hskip 1em plus 0.5em minus 0.4em\relax Oxford England: Butterworth-Heinemann, 1984.

\bibitem{schwinger1998classical}
J.~Schwinger~\textit{et al.}, \emph{Classical electrodynamics}.\hskip 1em plus 0.5em minus 0.4em\relax Reading, Mass: Perseus Books, 1998.

\bibitem{Wald2022electromagnetism}
R.~Wald, \emph{Advanced classical electromagnetism}.\hskip 1em plus 0.5em minus 0.4em\relax Princeton, NJ: Princeton University Press, Mar. 2022.

\bibitem{Kerns1976}
D.~M. Kerns, ``Plane-wave scattering-matrix theory of antennas and antenna-antenna interactions: formulation and applications,'' \emph{Journal of Research of the National Bureau of Standards, Section B: Mathematical Sciences}, vol. 80B, no.~1, p.~5, Jan 1976.

\bibitem{mikki_nf2}
S.~{Mikki} and Y.~M.~M. {Antar}, ``A theory of antenna electromagnetic near field -- {Part II},'' \emph{IEEE Transactions on Antennas and Propagation}, vol.~59, no.~12, pp. 4706--4724, Dec 2011.

\bibitem{rigorous_MC}
S.~Mikki and Y.~Antar, ``A rigorous approach to mutual coupling in general antenna systems through perturbation theory,'' \emph{IEEE Antennas and Wireless Communication Letters}, vol.~14, pp. 115--118, 2015.

\bibitem{mikki_energy}
S.~{Mikki} and Y.~M.~M. {Antar}, ``A new technique for the analysis of energy coupling and exchange in general antenna systems,'' \emph{IEEE Transactions on Antennas and Propagation}, vol.~63, no.~12, pp. 5536--5547, Dec 2015.

\bibitem{Kerns_1976}
D.~Kerns, ``{Plane Wave Scattering-Matrix Theory of Antennas and Antenna-Antenna Interactions},'' \emph{IEEE Antennas and Propagation Society Newsletter}, vol.~21, no.~1, pp. 11--11, February 1979.

\bibitem{10012689}
S.~Mikki, ``The shannon information capacity of an arbitrary radiating surface: An electromagnetic approach,'' \emph{IEEE Transactions on Antennas and Propagation}, vol.~71, no.~3, pp. 2556--2570, 2023.

\bibitem{10005192}
R.~Ji, S.~Chen, C.~Huang, J.~Yang, W.~E.~I. Sha, Z.~Zhang, C.~Yuen, and M.~Debbah, ``Extra dof of near-field holographic mimo communications leveraging evanescent waves,'' \emph{IEEE Wireless Communications Letters}, vol.~12, no.~4, pp. 580--584, 2023.

\bibitem{Dardari2020HMIMO}
D.~Dardari, ``Communicating with large intelligent surfaces: Fundamental limits and models,'' \emph{IEEE Journal on Selected Areas in Communications}, vol.~38, no.~11, pp. 2526--2537, 2020.

\end{thebibliography}

\end{document}